\documentstyle[12pt]{article}

\begin{document}

\hfill \vbox{ \hbox{UCLA/93/TEP/39}
              \hbox{hep-lat/9311005} }

\begin{center}

{\Large \bf The SU(2) $\times$ SU(2) chiral spin model in terms
of SO(3) and Z$_2$ variables: vortices and disorder}\footnote{Research
supported in part by NSF Grant PHY89-15286} \\[2cm]

{\bf Tam\'as G.\ Kov\'acs}\footnote{On leave of absence from the
Department of Theoretical
Physics, Kossuth Lajos University, Deb\-re\-cen, Hungary; present
e-mail address: kovacs@physics.ucla.edu}
 and {\bf E.\ T.\ Tomboulis}\footnote{e-mail address:
tomboulis@uclaph.bitnet} \\
{\em Department of Physics \\
University of California, Los Angeles \\
Los Angeles, California 90024-1547} \\[3.5cm]

\end{center}

\section*{Abstract}

We rewrite the two-dimensional SU(2)$\times$ SU(2)
chiral spin model in terms of SO(3) and {\bf Z}$_2$ degrees of
freedom. The transformation, which is motivated by a similar
representation of the corresponding lattice gauge theory in
higher dimensions, exhibits the presence of dynamical SO(3) vortices
and associated strings. We present arguments that (pairs of) SO(3)
vortices with long strings play a crucial role in disordering the
spin system at arbitrarily low temperatures.
\vfill

\pagebreak

Two dimensional spin models are well known to have properties analogous
to four dimensional gauge theories. For gauge theories on the lattice,
a convenient framework for a systematic comparison of the SU(N),
SU(N)/{\bf Z}$_N$ and mixed action models is provided by the
rewriting of the SU(N) theory in terms of SU(N)/{\bf Z}$_N$ and
{\bf Z}$_N$ variables. It has been known for some time that by means
of such a transformation the SU(N) theory can be represented as
a {\bf Z}$_N$ gauge theory coupled to dynamical SU(N)/{\bf Z}$_N$
monopole currents \cite{Tomboulis1}.
The effect of these monopoles and associated vortices
on the phase diagram as determined by bulk properties, as well as
on certain long-distance quantities such as the magnetic disorder parameter
('t Hooft loop) has been studied fairly extensively, mostly for
N=2, in a variety of situations \cite{Analgauge,Numgauge}. More
recently, this representation of the SU(2) theory has been
shown to be useful for addressing the problem of confinement
at arbitrarily weak coupling \cite{Tomboulis2}. It is well-known
that the 4d {\bf Z}$_2$ gauge theory has a weak coupling deconfined
phase, so we cannot expect the {\bf Z}$_2$ part of the SU(2)
model to produce confinement in itself. However, the coupling
to the SU(2)/{\bf Z}$_2 \cong$ SO(3) monopoles and their Dirac
sheets sufficiently disorders the {\bf Z}$_2$ system to avoid
a transition, provided monopole current correlations obey certain
bounds. Thus the confinement problem is reduced to estimates on
monopole expectations at large $\beta$.

The Monte Carlo study of 2d SU(N) and SU(N)/{\bf Z}$_N$ chiral
spin models suggests a physical picture completely analogous to
the results obtained for SU(N) gauge theories \cite{Kogut2}. In particular,
SU(N)/{\bf Z}$_N$ vortices, the analogues of monopoles, seem to
play an important role in disordering the system. In view of
these results, it is somewhat surprising that the explicit isolation
of the vortices and their couplings in the partition function
measure has not been performed before. In this paper we derive
the representation of the partition function and the 2pt correlation
function of the 2d SU(2) $\times$ SU(2) chiral spin model in terms
of {\bf Z}$_2$ and SO(3) variables, and explicitly exhibit the
vortices as part of the measure. Using this representation
we give arguments supporting the role of SO(3) vortices and
their strings (the analogs of monopoles and Dirac strings in
gauge theory) in disordering the system at large $\beta$. \\[2mm]

We shall work on a finite two-dimensional sqare lattice
$\Lambda$ with free boundary conditions. The elementary
constituents of the lattice are simplices of various
dimensions and will be denoted by $p$ (plaquettes), $l$
(links) and $s$ (sites). It will be convenient to use the
notations of (co)homology theory. In this language abelian
group valued configurations can be described by chains
(a k-chain is an assignement of group elements to all
k-simplices). The (exterior) differential and codifferential
operators are denoted by d and $\delta$ respectively.
The SU(2)$\times$SU(2) chiral spin model is defined by the
action
\begin{equation}
 {\cal A} = - \sum_{l \in \Lambda} \mbox{tr} U_l,
     \label{eq:action}
\end{equation}
where SU(2) elements $U_s$ are attached to lattice sites and
$U_l=U_s^\dagger U_{s'}$ with $[ss'] = \partial l$
($\partial l$ denotes the boundary of $l$). The model
is analogous to an SU(2) gauge theory, the only
difference being that in the gauge model each corresponding
object lives on simplices one dimension higher than in
the spin model. The analogy makes it possible to translate
the method described in \cite{Tomboulis1} to the language of
the spin model and rewrite the partition function in a
similar fashion.

The partition function  (PF) of the spin model is given by
\begin{equation}
 Z = \prod_{s \in \Lambda} \int \mbox{d}U_s
 \hspace{2mm} \exp \left[ \beta \sum_{l \in \Lambda}
 \mbox{tr} U_l \right],
\end{equation}
where d$U_s$ is the Haar-measure on SU(2) normalised to
unity and the product runs over all lattice sites.
Throughout this paper all group integrations will be
performed using the invariant measure normalised to
unity, regardless of the discrete or continuous nature
of the group. The invariance of the group measure
guarantees that a shift of the variables $U_s \rightarrow \gamma_s
U_s$, where $\gamma_s \in$ {\bf Z}$_2$ at each site does
not affect the PF. Since the PF does not depend on the $\gamma$
configuration, we can integrate on all these $\gamma$ variables.
Thus
\begin{equation}
 Z = \prod_{s \in \Lambda} \int \mbox{d}U_s
 \int \mbox{d} \gamma_s \hspace{2mm}
 \exp \left[ \beta \sum_{l \in \Lambda} (\mbox{d} \gamma)_l
 \mbox{tr} U_l \right],
\end{equation}
where
\begin{equation}
(\mbox{d} \gamma)_l = \prod_{s \in \partial l} \gamma_s
\end{equation}
Introducing the notations
\begin{equation}
 K(U_l) = \beta |\mbox{tr} U_{l}| \hspace{1cm} \mbox{and}
 \hspace{1cm} \eta_l = \mbox{sign} \; \mbox{tr} U_l,
     \label{eq:defeta}
\end{equation}
the PF can be written as
\begin{equation}
 Z = \prod_{s \in \Lambda} \int \mbox{d}U_s
 \int \mbox{d} \gamma_s \;
 \exp \left[ \sum_{l \in \Lambda} K(U_l) \eta_l
 (\mbox{d} \gamma)_l \right].
\end{equation}
The {\bf Z}$_2$-valued 1-chain, $\eta$ is determined by
the spin configuration $[U]$ through (\ref{eq:defeta}) and
$\eta_l=-1$ means that the bond $l$ is highly excited.
Now it is convenient to introduce a new {\bf Z}$_2$-valued
1-chain $\sigma_l$ with the definition
\begin{equation}
 \sigma_l = \eta_l \; (\mbox{d} \gamma)_l.
     \label{eq:defsigma}
\end{equation}
The constraint (\ref{eq:defsigma}) is enforced
by a delta function on each link, giving
\begin{equation}
 Z = \prod_{s \in \Lambda} \int \mbox{d}U_s
 \int \mbox{d}\gamma_s \prod_{l \in \Lambda}
 \int \mbox{d}\sigma_l \;
 \, \delta \left( \sigma_l^{-1} \eta_l (\mbox{d} \gamma)_l
 \right) \hspace{2mm} \exp \left[ \sum_{l \in \Lambda}
 K(U_l) \sigma_l \right].
\end{equation}
Since the $\gamma$ variables are contained only in the delta functions,
the $\gamma$-integrations can be carried out, resulting in the
constraint that the 1-chain $\sigma^{-1} \eta$ is closed.
In this way the PF is obtained in the following form
\begin{equation}
 Z = \prod_{s \in \Lambda} \int \mbox{d}U_s
 \prod_{l \in \Lambda} \int \mbox{d}\sigma_l
 \prod_{p \in \Lambda} \delta((\mbox{d} \sigma^{-1})_p
 (\mbox{d} \eta)_p) \; \exp \left[ \sum_{l \in \Lambda}
 K(U_l) \sigma_l \right].
     \label{eq:Ztrans}
\end{equation}
The remarkable property of this form of the PF is that
the integrand depends on the SU(2) degrees of freedom $U_s$ only through
the SU(2)/{\bf Z}$_2$ cosets.  In other words, it has a $U_s \rightarrow
-U_s$ ``gauge'' symmetry and effectively the spins can be regarded
SU(2)/{\bf Z}$_2 \cong$ SO(3) variables rather than SU(2) ones.  Of
course, the price that is paid for this extra symmetry is the appearance
of the new {\bf Z}$_2$ degrees of freedom ($\sigma_l$) attached to links
of the lattice.

There are two different couplings between the SO(3) and the {\bf Z}$_2$
degrees of freedom.  Firstly, $\sigma_l$ contributes an extra sign to
the coupling between the SO(3) spins
residing on the two ends of the link $l$.  Secondly, there is a coupling
through the delta function constraint.

The physical meaning of the quantities d$\sigma$ and d$\eta$
appearing in the constraint is the following. If the
$\sigma$-field is viewed as a {\bf Z}$_2$ gauge field, then
$\mbox{d} \sigma$ is the associated curvature (recall that
$(\mbox{d} \sigma)_p = \prod_{l \in \partial p} \sigma_l$).
$(\mbox{d} \eta)_p = -1$ means that there is an odd number of
``negative'' $(\eta_l = -1)$ links around $p$, i.e.\ the plaquette $p$
carries an SO(3) ``vortex''.  The ``charge'' of SO(3) vortices in two
dimensions is characterised by elements of {\bf Z}$_2$, the fundamental
group of SO(3).  Notice that in terms of the SO(3) variables, the location of
negative-$\eta$ links is ambiguous. They can be moved around by changing
the representative elements of cosets. On the other hand
the position of vortices is gauge invariant, because a
$U_s \rightarrow -U_s$ transformation flips the sign of two $\eta_l$-s
around the affected plaquettes.  It
is also clear from the construction that each SO(3) vortex is attached
to a string of negative-$\eta$ links which terminates in another
vortex or on the boundary of the lattice.
These strings can be deformed by changing coset representatives but
their end-points, the vortices are fixed.  The situation is analogous to
gauge theories, where one has Dirac-strings attached to monopoles and
these strings can be deformed by gauge transformations.  The
delta function in (\ref{eq:Ztrans}) constrains the $\sigma$-curvature to
be equal to the SO(3) vortex number on each plaquette.
In this way vortices are also connected by strings of negative
$\sigma$ links (Fig.\ 1).

It should be noted that there is a substantial difference between
$\sigma$ and $\eta$ strings. To see this, let us look at a closed
contour of links enclosing exactly one vortex. Both the $\sigma$ and
the $\eta$ string attached to the vortex have to pierce the
contour somewhere. The link $l$, where the $\sigma$ string intersects
the contour is unambiguously given by the $[\sigma]$
configuration and $l$ carries an energy $2K(U_l)$.
On the other hand, the location where the $\eta$ string crosses
the contour ($l'$) has no physical meaning in terms of the
SO(3) variables; it can be moved by choosing different representative
elements of the cosets at some sites. It follows that the energy
of the $\eta$ string is not necessarily localised on the
$\eta_l=-1$ links. Indeed, for most of the configurations the
SO(3) spins change slowly along the contour. This is more favourable
both in terms of energy and entropy, than having an abrupt
change somewhere. The analogous mechanism in lattice gauge
theories is called flux spreading, and it is believed to play
an important role in confinement \cite{Yaffe,Analgauge}. It can be seen
from the above argument that the energy cost of a $\sigma$
string is necessarily proportional to its length. On the other hand,
many SO(3) configurations can be produced for which
the energy cost of an $\eta$ string is constant, independent of
its length. As a consequence, at low temperatures
the $\sigma$ string connecting two given vortices is very likely
to be of minimal length, but the corresponding $\eta$ string can fluctuate
considerably (Cp.\ Fig.\ 1).

It is now instructive to perform a duality transformation on the
{\bf Z}$_2$ degrees of freedom, which amounts to trading the
$\sigma_l$ link variables for plaquette variables
$\omega_p \in$ {\bf Z}$_2$. In this way the expression of
the PF becomes similar to that of an Ising
model on the dual lattice with fluctuating couplings. Technically
the duality transformation is done by expanding the
Boltzmann weights in (\ref{eq:Ztrans}) in characters of
{\bf Z}$_2$ and carrying out the $\sigma$ integrations.
The dual form of the PF is then obtained in the form
\begin{equation}
 Z = \prod_{s \in \Lambda} \int \mbox{d}U_s \,
 \exp \left( \sum_{l \in \Lambda} \hat{M}(U_l) \right)
 \prod_{p \in \Lambda} \int \mbox{d}\omega_p
 \chi_{(d \eta)_p}(\omega_p) \exp \left[ \sum_{l \in \Lambda}
 \hat{K}(U_l) (\delta \omega)_l \right],
     \label{eq:Zomega}
\end{equation}
where
\begin{equation}
 \hat{K}(U_l) = \frac{1}{2} \ln \coth K(U_l), \hspace{6mm}
 \hat{M}(U_l) = \frac{1}{2} \ln \left( \cosh K(U_l) \sinh K(U_l)
 \right),
\end{equation}
$(\delta \omega)_l = \prod_{p \owns l} \omega_p$ and
$\chi_q$ are the characters of {\bf Z}$_2$. This form of the partition
function contains {\bf Z}$_2$ spins ($\omega_p$), attached
to plaquettes. Spins on the two plaquettes sharing the link $l$
interact via the fluctuating coupling $K(U_l)$. The group characters
couple these {\bf Z}$_2$ spins to SO(3) vortices.

Let us now consider the 2pt correlation function. We follow the
same procedure as in the case of the PF, but now take periodic
b.c.\ to within elements of {\bf Z}$_2$, i.e.\ periodic b.c.\
for the coset variables. Then the expectation $\langle \mbox{tr}
(U^\dagger_x U_{x'}) \rangle$ in terms of the new variables is
\begin{eqnarray}
 Z  \; \langle \mbox{tr} (U_x^\dagger U_{x'}) \rangle =
  \prod_{s \in \Lambda} \int \mbox{d}U_s \,
 \exp \left( \sum_{l \in \Lambda} \hat{M}(U_l) \right)
     \nonumber \\
 \times \prod_{p \in \Lambda} \int \mbox{d}\omega_p
 \chi_{(d \eta)_p}(\omega_p) \eta_C \, \mbox{tr}(U_x^\dagger U_{x'}) \,
 \exp \left[ \sum_{l \in \Lambda}
 \hat{K}(U_l) (\delta \omega)_l (-1)^{E(C)_l} \right],
     \label{eq:2pt}
\end{eqnarray}
where $C$ is an arbitrary path of links connecting the sites
$x$ and $x'$, $\eta_C= \prod_{l \in C} \eta_l$ and $E(C)_l$ is the
characteristic function of $C$, i.e.\ it is 1 if $l \in C$,
0 otherwise. In the r.h.s.\ of (\ref{eq:2pt}) the $U_s$'s now obey periodic,
the $\omega_p$'s free b.c. A deformation of the path $C$ is equivalent to an
irrelevant shift of the $\omega$ integration variables, so the
correlation function is independent of $C$. It will be convenient
to choose $x$ and $x'$ on opposite edges of the lattice along
the ``1'' direction (Fig.\ 2a-b). Then by the periodic b.c.\ the
tr($U^\dagger_x U_{x'}$) factor on the r.h.s.\ reduces to unity and
(\ref{eq:2pt}) becomes the 2d spin analog of the familiar electric
flux free energy order parameter of gauge theories.

Let us now choose some fixed ``background'' configuration of
the SO(3) variables and integrate out all the {\bf Z}$_2$ degrees
of freedom. To understand the behaviour of the 2pt function,
we have to describe those SO(3) configurations that can give
considerably different contribution to the partition
function with a tr$(U_x^\dagger U_{x'})$ insertion, eqn.\ (\ref{eq:2pt}),
than without it, eqn.\ (\ref{eq:Zomega}). For a fixed $[U]$ configuration
there are two places where the two integrands differ:

\begin{itemize}

\item{There is a $(-1)^{E(C)_l}$ ``twist'' along the curve $C$,
that changes the sign of the couplings between spins on different
sides of $C$}

\item{$\eta_C$ measures the number (mod 2) of $\eta$ strings
piercing $C$}

\end{itemize}

The effect of the twist can be compensated by flipping all the
spins on one side (say above) of $C$. In other words, for a
given configuration $[\omega]$ there is another $[\tilde{\omega}]$
such that the exponent without the twist evaluated at $[\omega]$
is equal to the exponent with the twist at $[\tilde{\omega}]$.
Since the {\bf Z}$_2$ spin measure is invariant, SO(3)
configurations with no strings piercing
$C$ and no vortices give the same contribution to (\ref{eq:Zomega})
and (\ref{eq:2pt}).

Now let us look at an SO(3) configuration that contains
$V$ vortices above $C$ and $S$ string crossing points on $C$.
Upon changing from $[\omega]$ to $[\tilde{\omega}]$ each
vortex above C contributes an extra -1 factor (because of
the $\chi_{(d \eta)_p}(\omega_p)$ factors). Due to the
$\eta_C$ insertion, each string crossing point on $C$ gives
an additional minus sign. The net effect will be a relative
$(-1)^{V+S}$ sign between the contribution of $[\omega]$ to
(\ref{eq:Zomega}) and that of $[\tilde{\omega}]$ to
(\ref{eq:2pt}), the moduli of these contributions, upon performing
the {\bf Z}$_2$ integrations, being equal.
Notice that with an odd total number of vortices,
$\prod_{p \in \Lambda} \chi_{(d \eta)_p}(\omega_p)$ is odd
with respect to the global {\bf Z}$_2$ symmetry $\omega \rightarrow
-\omega$ of the rest of the integrand. This property ensures
that SO(3) configurations with an odd number of vortices
give vanishing contribution to both the PF and the
correlation function. Hence the number of vortices (mod 2)
above and below $C$ are the same and the above
argument remains valid with $V$ being the number of
vortices below $C$. Notice also that $(-1)^{V+S}$ is not affected
by any deformation of $C$. Whenever we lose or gain a vortex
above $C$ by a deformation of the path we also lose or gain an $\eta$
string crossing point on $C$.

We have seen that the important SO(3) configurations,
the ones that ``see'' the two-point insertion in the partition function,
are those with $V+S$ odd. There are two different types of
configurations that can give an odd $V+S$:

\begin{itemize}

\item{An $\eta$ string running all the way through the lattice
in the direction perpendicular to $C$ (Fig.\ 2a).}

\item{A pair of vortices residing on different sides of
$C$ with their strings terminating on the edge of the lattice
without piercing $C$ (Fig.\ 2b).}

\end{itemize}

The asymptotic behaviour of the correlation function for large
lattices depends upon the relative weight of these SO(3) configurations.
Pairs of vortices as in Fig.\ 2b incur an energy cost proportional
to their separation, because in (\ref{eq:2pt}) $\omega$ variables must
be excited in the intervening gap to obtain a nonzero contribution;
in the original variables, this reflects the necessary presence of
a $\sigma$ string. (Even in the SO(3) $\times$ SO(3) spin model, where
in (\ref{eq:action}) the adjoint representation character is used and
there is no $\sigma$ string, the energy cost grows as the log of the
separartion.) Thus pairs of vortices tend to cluster together at
large $\beta$. The crucial point, however, is that the configurations
of Fig.\ 2 contain long strings and the energetics of $\eta$ strings
is dominated by the SO(3) part $(\hat{M}(U))$ of the action in
(\ref{eq:2pt}). Simple semiclassical estimates indicate that
if $d \leq 2$, due to flux spreading,
the free energy cost of these long $\eta$ strings can remain finite in
the large lattice limit, at large $\beta$. If this is
the case, then the direct coupling of such $\eta$-strings to the
correlation function noted above, can disorder it, even if the density
of vortices the strings are attached to becomes exponentially small
($\sim \exp(-const. \beta)$) at large $\beta$. The resulting mass gap
will itself be exponentially small.

The remarkable effectiveness of SO(3) vortices in reducing the
correlation length by many orders of magnitude was indeed noted
in numerical simulations at large values of $\beta$, at which
asymptotically free perturbation theory predicts enormous
correlation lengths, while the vortices tend to cluster together
at exponentially small densities \cite{Kogut2}. The picture
presented here provides an explanation for this phenomenon.

It is clearly worthwile to substantiate this picture by rigorous
estimates. This can be approached along the lines suggested in
Ref. \cite{Tomboulis2} for the closely analogous case of SO(3)
monopoles and Dirac sheets in 4d. We will report on such estimates
elsewhere.

\section*{Figure captions}

\begin{tabular}{r p{11cm}}

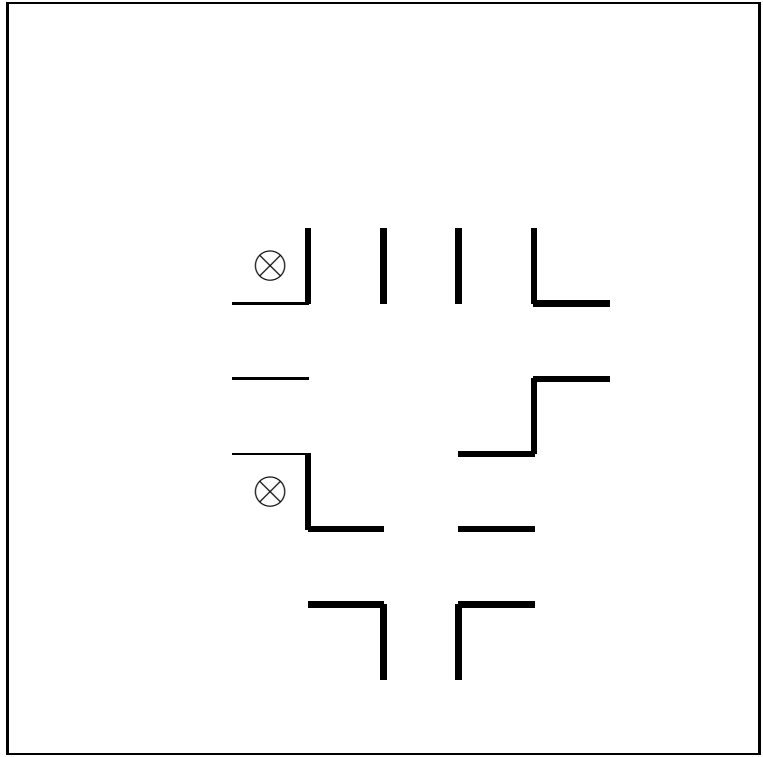
{\em Figure 1} & Two vortices connected by a $\sigma=-1$ and
an $\eta=-1$ string. \\[2mm]

{\em Figure2a} & An $\eta=-1$ string running all the way through
the lattice intersecting the path $C$. $C$ is an arbitrary path
connecting the points $x$ and $x'$. \\[2mm]

{\em Figure 2b} & A pair of vortices residing on different sides
of $C$ with their attached strings terminating on the edge of
the lattice.

\end{tabular}

\pagebreak

\unitlength=1.00mm
\linethickness{0.4pt}
\begin{picture}(110.00,160.00)
\put(10.00,30.00){\line(1,0){100.00}}
\put(110.00,30.00){\line(0,1){100.00}}
\put(110.00,130.00){\line(-1,0){100.00}}
\put(10.00,130.00){\line(0,-1){100.00}}
\put(40.00,70.00){\line(1,0){10.00}}
\put(50.00,80.00){\line(-1,0){10.00}}
\put(40.00,90.00){\line(1,0){10.00}}
\put(45.00,95.00){\makebox(0,0)[cc]{\Large $\otimes$}}
\linethickness{2.0pt}
\put(60.00,40.00){\line(0,1){10.00}}
\put(70.00,50.00){\line(0,-1){10.00}}
\put(70.00,50.00){\line(1,0){10.00}}
\put(80.00,60.00){\line(-1,0){10.00}}
\put(70.00,70.00){\line(1,0){10.00}}
\put(80.00,70.00){\line(0,1){10.00}}
\put(80.00,80.00){\line(1,0){10.00}}
\put(90.00,90.00){\line(-1,0){10.00}}
\put(80.00,90.00){\line(0,1){10.00}}
\put(70.00,100.00){\line(0,-1){10.00}}
\put(60.00,90.00){\line(0,1){10.00}}
\put(50.00,100.00){\line(0,-1){10.00}}
\put(10.00,10.00){\makebox(0,0)[lc]{\Large $\otimes$ vortex}}
\linethickness{0.4pt}
\put(40.00,10.00){\line(1,0){10.00}}
\put(54.00,10.00){\makebox(0,0)[lc]{\Large $\sigma_l=-1$}}
\linethickness{2.0pt}
\put(85.00,10.00){\line(1,0){10.00}}
\put(99.00,10.00){\makebox(0,0)[lc]{\Large $\eta_l=-1$}}
\put(0.00,160.00){\makebox(0,0)[lc]{{\LARGE \bf Figure 1}}}
\put(50.00,70.00){\line(0,-1){10.00}}
\put(50.00,60.00){\line(1,0){10.00}}
\put(60.00,50.00){\line(-1,0){10.00}}
\put(45.00,65.00){\makebox(0,0)[cc]{{\Large $\otimes$}}}
\end{picture}

\pagebreak

\unitlength=1.00mm
\linethickness{0.4pt}
\begin{picture}(114.00,159.00)
\put(10.00,30.00){\line(1,0){100.00}}
\put(110.00,30.00){\line(0,1){100.00}}
\put(110.00,130.00){\line(-1,0){100.00}}
\put(10.00,130.00){\line(0,-1){100.00}}
\linethickness{2.0pt}
\put(55.00,10.00){\line(1,0){10.00}}
\put(69.00,10.00){\makebox(0,0)[lc]{\Large $\eta_l=-1$}}
\put(0.00,159.00){\makebox(0,0)[lc]{{\LARGE \bf Figure 2a}}}
\put(50.00,40.00){\line(1,0){10.00}}
\put(60.00,50.00){\line(-1,0){10.00}}
\put(50.00,60.00){\line(1,0){10.00}}
\put(60.00,70.00){\line(-1,0){10.00}}
\put(50.00,80.00){\line(1,0){10.00}}
\put(60.00,90.00){\line(-1,0){10.00}}
\put(50.00,100.00){\line(1,0){10.00}}
\put(60.00,110.00){\line(-1,0){10.00}}
\put(50.00,120.00){\line(1,0){10.00}}
\put(60.00,130.00){\line(-1,0){10.00}}
\put(50.00,30.00){\line(1,0){10.00}}
\linethickness{0.4pt}
\put(10.00,70.00){\line(1,0){100.00}}
\put(114.00,70.00){\makebox(0,0)[lc]{{\Large $x'$}}}
\put(6.00,70.00){\makebox(0,0)[rc]{{\Large $x$}}}
\put(90.00,74.00){\makebox(0,0)[cb]{{\Large $C$}}}
\end{picture}
\pagebreak

\unitlength=1.00mm
\linethickness{0.4pt}
\begin{picture}(114.00,160.00)
\put(10.00,30.00){\line(1,0){100.00}}
\put(110.00,30.00){\line(0,1){100.00}}
\put(110.00,130.00){\line(-1,0){100.00}}
\put(10.00,130.00){\line(0,-1){100.00}}
\put(20.00,10.00){\makebox(0,0)[lc]{\Large $\otimes$ vortex}}
\linethickness{2.0pt}
\put(75.00,10.00){\line(1,0){10.00}}
\put(89.00,10.00){\makebox(0,0)[lc]{\Large $\eta_l=-1$}}
\linethickness{0.4pt}
\put(0.00,160.00){\makebox(0,0)[lc]{{\LARGE \bf Figure 2b}}}
\put(10.00,70.00){\line(1,0){100.00}}
\linethickness{2.0pt}
\put(70.00,30.00){\line(1,0){10.00}}
\put(80.00,40.00){\line(-1,0){10.00}}
\put(70.00,50.00){\line(1,0){10.00}}
\put(50.00,80.00){\line(1,0){10.00}}
\put(50.00,90.00){\line(1,0){10.00}}
\put(60.00,90.00){\line(0,1){10.00}}
\put(60.00,100.00){\line(1,0){10.00}}
\put(70.00,110.00){\line(-1,0){10.00}}
\put(60.00,120.00){\line(1,0){10.00}}
\put(70.00,130.00){\line(-1,0){10.00}}
\put(114.00,70.00){\makebox(0,0)[lc]{{\Large $x'$}}}
\put(90.00,74.00){\makebox(0,0)[cb]{{\Large $C$}}}
\put(6.00,70.00){\makebox(0,0)[rc]{{\Large $x$}}}
\put(55.00,75.00){\makebox(0,0)[cc]{{\Large $\otimes$}}}
\put(75.00,55.00){\makebox(0,0)[cc]{{\Large $\otimes$}}}
\end{picture}

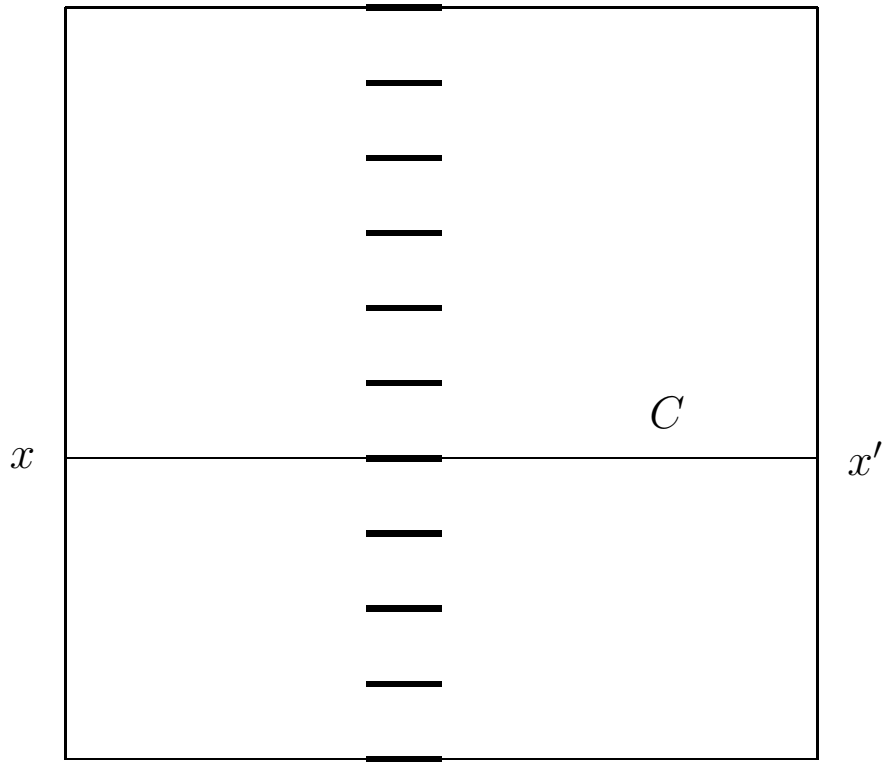
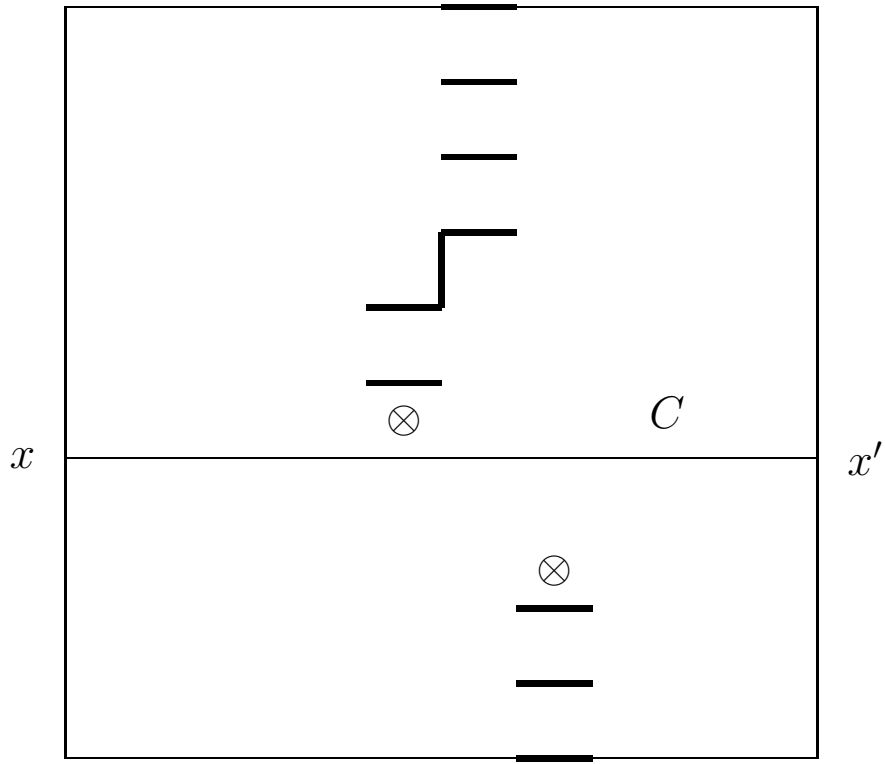
\end{document}